\documentstyle[nato,numreferences,epsf,amsfonts]{crckapb} 

\begin{opening}
\title{LOW-LYING DIRAC EIGENMODES, TOPOLOGICAL CHARGE\protect\\
       FLUCTUATIONS AND THE INSTANTON LIQUID MODEL}

\author{I.~HORV\'ATH, S.J.~DONG, T.~DRAPER}
\institute{Department of Physics \& Astronomy\\
           University of Kentucky, Lexington, KY 40503, USA}
\author{F.X.~Lee}
\institute{Center for Nuclear Studies and Department of Physics\\
           George Washington University\\
           Washington, DC 20052, USA, and\\
           Jefferson Laboratory, Newport News, VA 23606, USA}
\author{H.B.~Thacker}
\institute{Department of Physics \\
           University of Virginia, Charlottesville, VA 22901, USA}
\author{J.B.~Zhang}
\institute{CSSM and Department of Physics and Mathematical Physics \\
           University of Adelaide, Adelaide, SA 5005, Australia}           

\end{opening}

\begin{document}


%
%
\begin{abstract}
The local structure of low-lying eigenmodes of the overlap Dirac operator
is studied. It is found that these modes cannot be described as linear 
combinations of 't Hooft ``would-be'' zeromodes associated with instanton 
excitations that underly the Instanton Liquid Model. This implies 
that the instanton liquid scenario for spontaneous chiral symmetry breaking 
in QCD is not accurate. More generally, our data suggests that the vacuum 
fluctuations of topological charge are not effectively dominated by localized 
lumps of unit charge with which the topological ``would-be'' zeromodes could 
be associated.
\end{abstract}
\vspace*{-0.7cm}
%
%
\renewcommand{\thefootnote}{\fnsymbol{footnote}}
\footnotetext[0]{Presented by I.~Horv\'ath at the NATO Advanced Research 
Workshop ``Confinement, Topology, and other Non-Perturbative Aspects of 
QCD'', January 21-27, 2002, Star\'a Lesn\'a (Slovakia).}
\renewcommand{\thefootnote}{\arabic{footnote}}

\section{Introduction}

The idea of the instanton-dominated QCD vacuum appeared shortly after the discovery 
of the instanton~\cite{ins_disc}. In particular, the instanton gas 
picture~\cite{ins_gas} populates the vacuum with well-separated instantons which, 
using semiclassical methods, leads to the possible qualitative resolution of the $U(1)$ 
problem~\cite{ins_tHooft}, as well as to the conclusion that QCD physics depends
on the $\theta$-parameter~\cite{ins_theta}. However, it became clear very soon that 
QCD is not semiclassical in this sense because the instanton gas picture does not lead 
to confinement. In fact, it has been argued by Witten~\cite{WittenUA(1)} that large 
quantum fluctuations entirely destroy the semiclassical vacuum and that individual 
instantons do not play a significant dynamical role in QCD. 

The fundamental importance of instantons thus became clearly questionable, but the
instanton solution served as a basis for a phenomenological Instanton Liquid Model 
(ILM)~\cite{ILM}. While there are correlations among ILM instantons, the ILM vacuum 
is still a dilute medium where the instantons of size $\rho\approx 1/3$ fm and
density $n=1$ fm$^{-4}$ preserve their identity. It is thus possible to associate 
a 't Hooft ``would-be'' zeromode with an individual instanton. The mixing of these 
modes and a formation of ``topological subspace'' of low-lying modes is a basis for 
an elegant mixing scenario for spontaneous chiral symmetry breaking 
(S$\chi$SB)~\cite{ins_mixing}. Our goal will be to examine whether this effective 
picture can be recognized in the low-lying Dirac modes and thus whether one should 
assign a fundamental significance to it. The main conclusion from this study is 
that this is not the case~\cite{Hor02A} and that the ILM picture is not microscopically 
accurate.

\begin{figure}
\centering
\epsfxsize=8.0cm
\epsfysize=8.0cm
\centerline{\epsffile{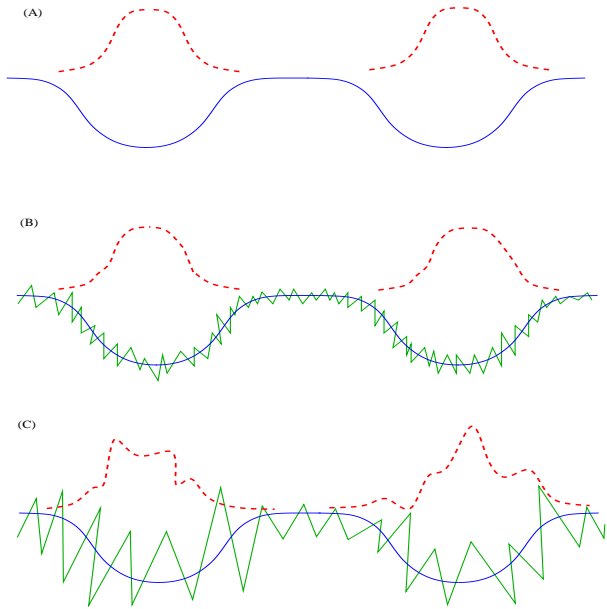}}
\caption{Short-distance fluctuations (rough line) imposed on the instanton-like
         gauge potential (smooth solid line) and its possible effects on 't Hooft
         modes (dashed line). See discussion in the text.}
\vspace{-0.0cm}
\end{figure}

Our approach is entirely based on studying the {\it local} properties of the 
fermionic eigenmodes~\cite{Hor02A,Hor01A}. This is entirely appropriate if one's goal 
is to study the effects of topological charge fluctuations on S$\chi$SB. The underlying 
rough gauge fields are not manipulated in any way, while the fermionic dynamics generates 
the appropriately smooth behavior automatically. Consequently, this represents 
an {\it unbiased} approach to study the possible dynamical relevance of ILM instantons, 
as well as the applicability of the topological mixing scenario in general.

The nature of the questions we ask can be described by starting from the ILM picture 
of the vacuum, or from any picture with vacuum populated by individual instantons 
preserving their identity. In a typical configuration, there is a collection of smooth 
(anti)self-dual potential wells for (right)left-handed components of the fermion. 
In Fig.~1(A) we show schematically the section of the instanton potential in a direction 
of another instanton. The potential is truly instanton-like around the core with 
modifications close to edges due to interactions. The corresponding left-handed 't Hooft 
modes are shown as well. In equilibrium QCD configuration there are certainly additional 
short-distance fluctuations present. However, these do not necessarily affect the 
low-momentum propagation of light quarks significantly. Indeed, the infrared Dirac modes
filter out the ultraviolet fluctuations that are not important for low-energy fermion 
dynamics~\cite{Hor02A}. It would thus be interesting to determine which of the following 
physically distinct possibilities takes place: (I) The strength and nature of quantum 
fluctuations is such that the structure of t' Hooft ``would-be'' zeromodes can still
be identified in the true eigenmodes as represented in Fig.~1(B). In this case the ILM 
scenario for S$\chi$SB could indeed be microscopically accurate. (II) Quantum fluctuations 
deform the 't Hooft modes to the extent that their original structure can not be recognized 
anymore (see Fig.~1(C)), but the unit quantization of topological charge still takes place 
in the QCD vacuum. In this case the ILM scenario would not be accurate and it is questionable 
whether the underlying gauge structures should be referred to as ``instantons''. However, 
as pointed out in Ref.~\cite{Hor02A}, the subspace of topological modes would nevertheless 
be created and thus the basic mechanism proposed in the context of ILM would be operating. 
(III) Quantum fluctuations are so strong that they destroy unit lumps. In this case 
a new microscopic origin of Dirac near-zeromodes needs to be sought. 

We will present data which rules out option (I) and strongly suggests that (II) is not 
applicable either. Since this talk has been given, the possibility (II) has been excluded 
directly~\cite{Hor02B}, leaving us with option (III).

\section{Structures in the Eigenmodes of the Overlap Operator}

The microscopic viability of the ILM picture for S$\chi$SB can in principle be verified
on the lattice by inspecting the local structure of low-lying Dirac modes~\cite{Hor01A}. 
In the absence of exact chiral symmetry this should be approached with some care, but
since chirally symmetric fermionic actions are now available~\cite{Neu98BA}, such  
strategy is very appropriate. We have calculated low-lying eigenmodes of the overlap 
Dirac operator~\cite{Neu98BA} in Wilson gauge backgrounds over a large span of lattice
spacings as summarized in Table~1. The physical volumes are chosen to contain
on average 3--4 instantons and antiinstantons if the ILM scenario is relevant, ensuring 
that the mixing of 't Hooft modes would take place, and that ``topological subspace''
of low-lying modes would form. If that happened, then the local structure of these modes 
would have very specific local properties related to the structure of underlying 't Hooft 
modes themselves. The logic of our approach is to {\it assume} that this is indeed 
the case, and to verify the consistency of such assumption against the {\it true} 
local behavior of low-lying modes. To do that, 
we identify the ``structures'' in the low-lying modes in such a way that (a) they would 
correspond to individual 't Hooft modes if ILM scenario was accurate and that (b) they 
can not arise accidentally as an artifact of the chosen procedure. 

\begin{table}[t]
  \centering
  \begin{tabular}{ccccccc}
  \hline\hline
  \multicolumn{1}{c}{$\beta$}  &
  \multicolumn{1}{c}{$\quad$}  &
  \multicolumn{1}{c}{$a$ [fm]}  &
  \multicolumn{1}{c}{$\quad$}  &
  \multicolumn{1}{c}{$V$}  &
  \multicolumn{1}{c}{$\quad$}  &
  \multicolumn{1}{c}{\# configs} \\
  \hline
  5.85 & & 0.123 & & $10^3$x$20$ & & 12 \\
  6.00 & & 0.093 & & $14^4$      & & 12 \\
  6.20 & & 0.068 & & $20^4$      & & 8  \\
  6.55 & & 0.042 & & $32^4$      & & 5 \\
\hline \hline
\end{tabular}
\caption{Ensembles of Wilson gauge configurations.}
\label{ensemb_tab} 
\end{table}

As a first step toward defining individual structures correspondingly, we study 
the $X$-distributions in lowest-lying near-zeromodes~\cite{Hor01A}. For a given 
eigenmode $\psi(n)$ the local chiral orientation parameter $X(n)$ is defined as
\begin{equation}
    \tan\,\left(\frac{\pi}{4}(1+X) \right) = \frac{|\psi_L|}{|\psi_R|} \;,
    \label{eq:10}
\end{equation}
where $\psi_L$, $\psi_R$ are the left and right spinorial components. $X(n)$ represents 
an angle in the $|\psi_L|$-$|\psi_R|$ plane rescaled so that $X(n)=-1$ for a purely 
right-handed spinor and $X(n)=+1$ for a purely left-handed one. For modes in the topological 
subspace the probability distribution of $X$ over the subvolumes occupied by (anti)instantons 
should be strongly peaked around $\pm 1$. On the other hand, if in the subvolume with 
strong fields the corresponding spinors were strictly unpolarized, a uniform 
distribution would result. 

\begin{figure}
\begin{center}
\centerline{
\epsfxsize=6.5truecm\epsffile{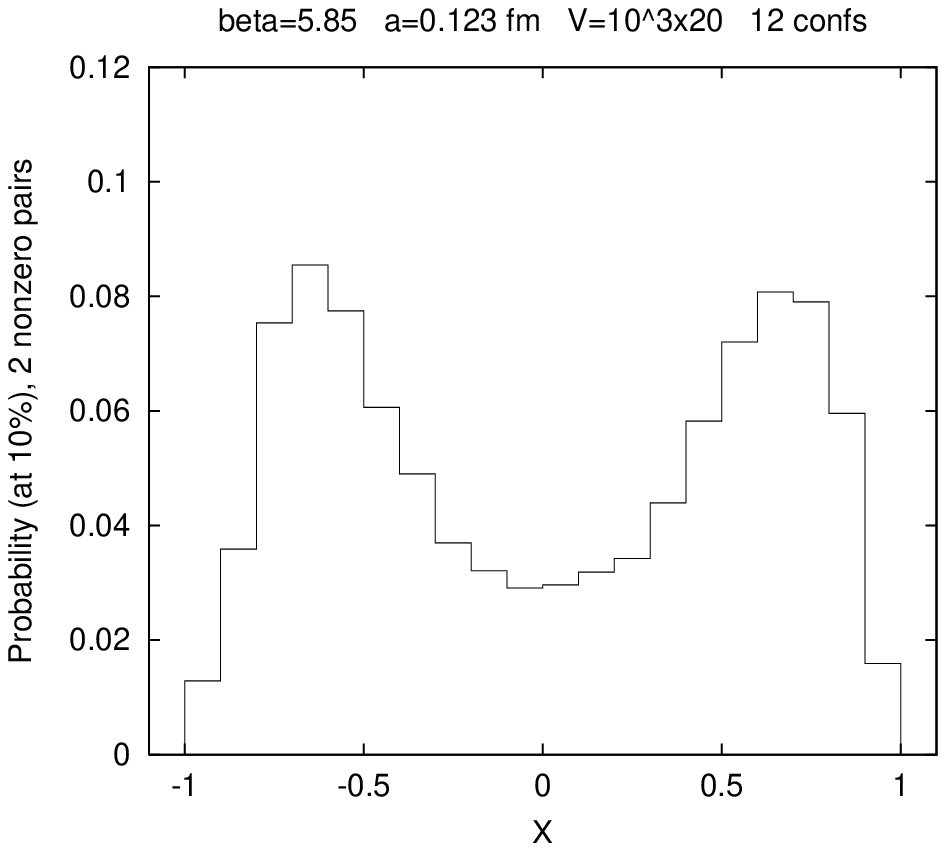}
\hspace*{-0.2cm}
\epsfxsize=6.5truecm\epsffile{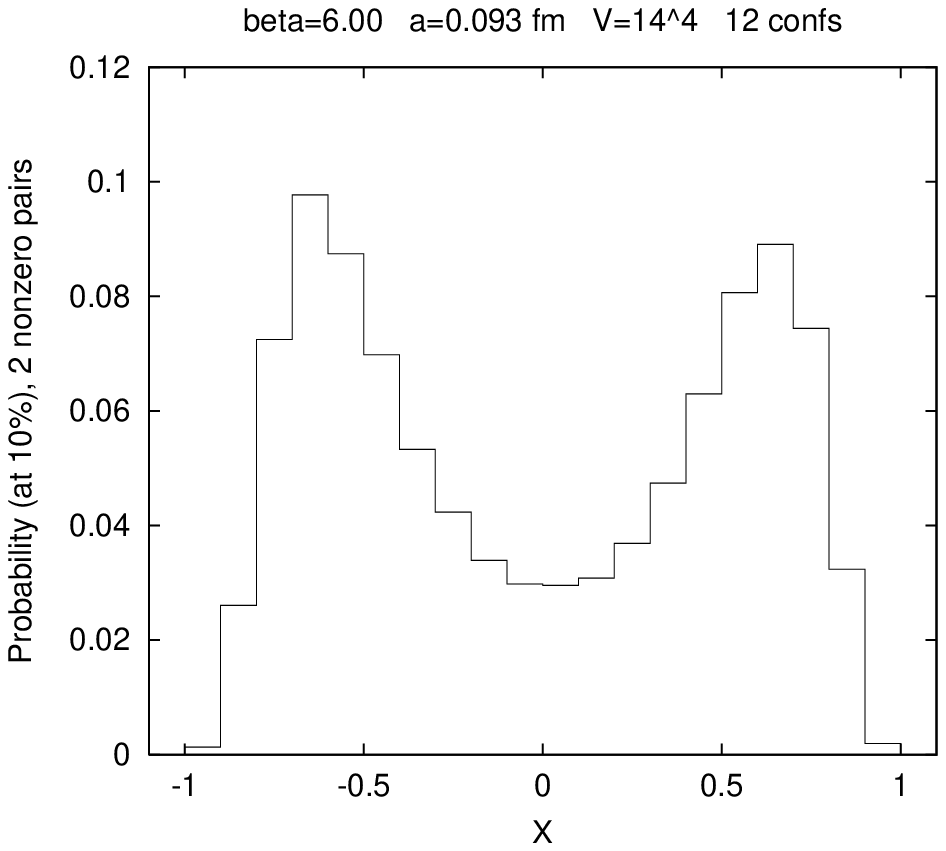}
}
\vskip 0.15in
\centerline{
\epsfxsize=6.5truecm\epsffile{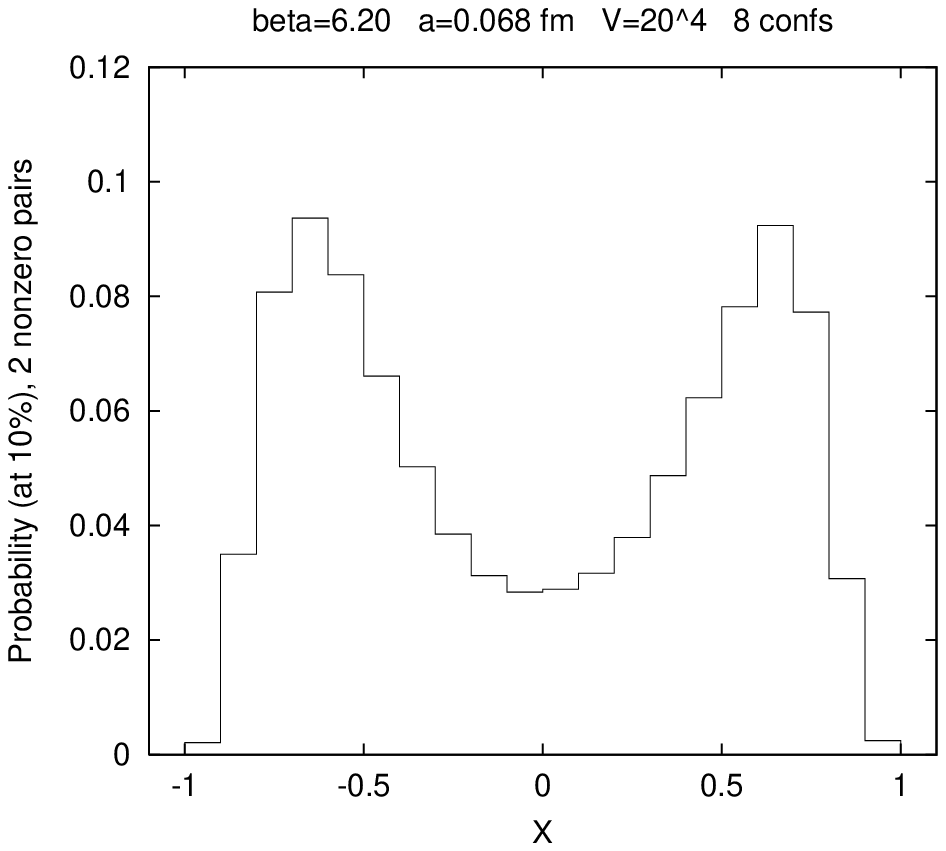}
\hspace*{-0.2cm}
\epsfxsize=6.5truecm\epsffile{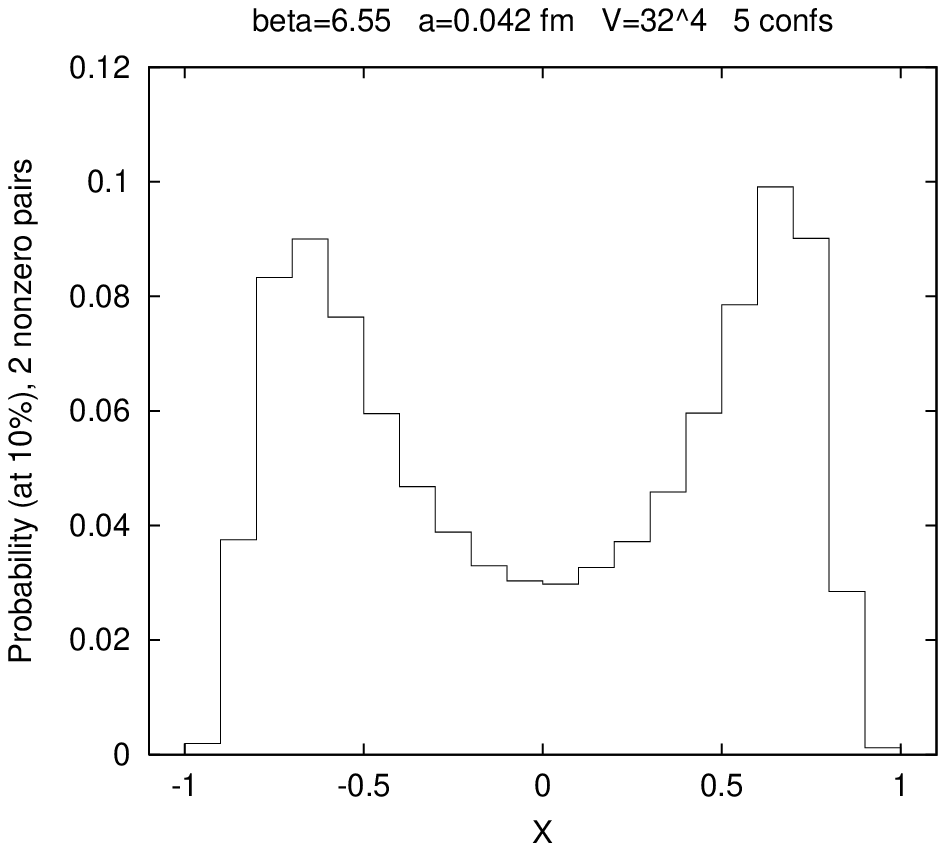}
}
\caption{$X$-distributions for four Wilson gauge ensembles considered.}
\label{Xdists}
\end{center}
\vskip -0.5cm
\end{figure}

We have calculated the lowest two pairs of near-zeromodes for the ensembles in Table~1.
The corresponding $X$-distributions shown in Fig.~\ref{Xdists} were obtained by 
considering the fraction $f=1/10$ of the volume, where the density 
$d(n)=\psi_n^{+}\psi_n$ is highest. This is plausible since the ILM estimates 
of the instanton packing fraction range between $f\approx1/20$ to $f\approx 1/8$. 
Also, the results are quite insensitive to variations of $f$ in the above range.
The $X$-distributions scale well, and a double-peaked behavior with maxima
at $X \approx \pm 0.65$ emerges (see also Refs.~\cite{Followup}). As 
Fig.~\ref{Xdists} shows, we do not obtain convincing peaks at around $\pm 1$ as one 
would naively expect from ILM (The calculated distributions for ILM ensembles are 
not available in the literature.). However, we use these $X$-distributions as 
a starting point to verify whether the double-peak behavior is due to the existence
of local structures resembling the ILM instantons. 

\begin{figure}
\begin{center}
\centerline{
\epsfxsize=9.0truecm\epsffile{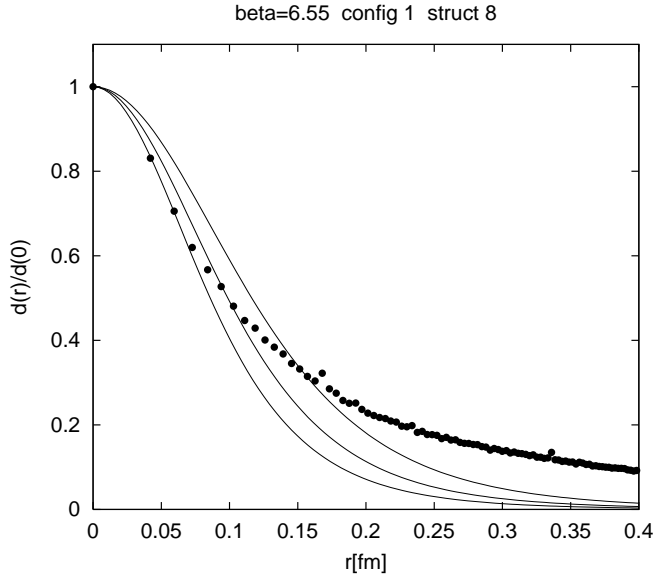}
}
\caption{The profile of the typical structure and attempted fits to 't Hooft
profile in the region
$0.00-0.06\,\mbox{\rm fm}$ (leftmost curve), 
$0.06-0.12\,\mbox{\rm fm}$ (middle curve), and 
$0.12-0.18\,\mbox{\rm fm}$ (rightmost curve).}
\label{Instfit}
\end{center}
\vskip -0.6truecm
\end{figure}

Following the above considerations we identify the local ``structure'' in the eigenmode 
by specifying the position $n$ of a local maximum of density $d(m)$ over the distance 
$\sqrt{3}$, such that (i) $|X(n)|\ge 0.5$, (ii) $n$ belongs to the subvolume used 
for calculating the $X$-distribution, and (iii) the density decays on average over
the distance $\sqrt{3}$ in eight basic lattice directions~\cite{Hor02A}. This 
definition satisfies the criteria (a),(b) discussed above and the resulting structures 
contribute to the maxima of the $X$-distribution by construction. To see whether 
a structure positioned at $n$ 
resembles the 't Hooft mode, we have attempted to fit the average profile of the
density $d_n(r) \,\equiv\, < d(m) >_{|n-m|=r}$ with the 't Hooft profile. The
situation for a typical structure is shown in Fig.~\ref{Instfit}. As can be seen, 
the fits over different regions are very poor and inconsistent with one another. 
We were simply unable to identify space-time structures whose origin could
be directly traced to 't Hooft zeromodes.

\section{Sizes and Densities of the Structures}

\begin{figure}
\begin{center}
\centerline{
\epsfxsize=9.0truecm\epsffile{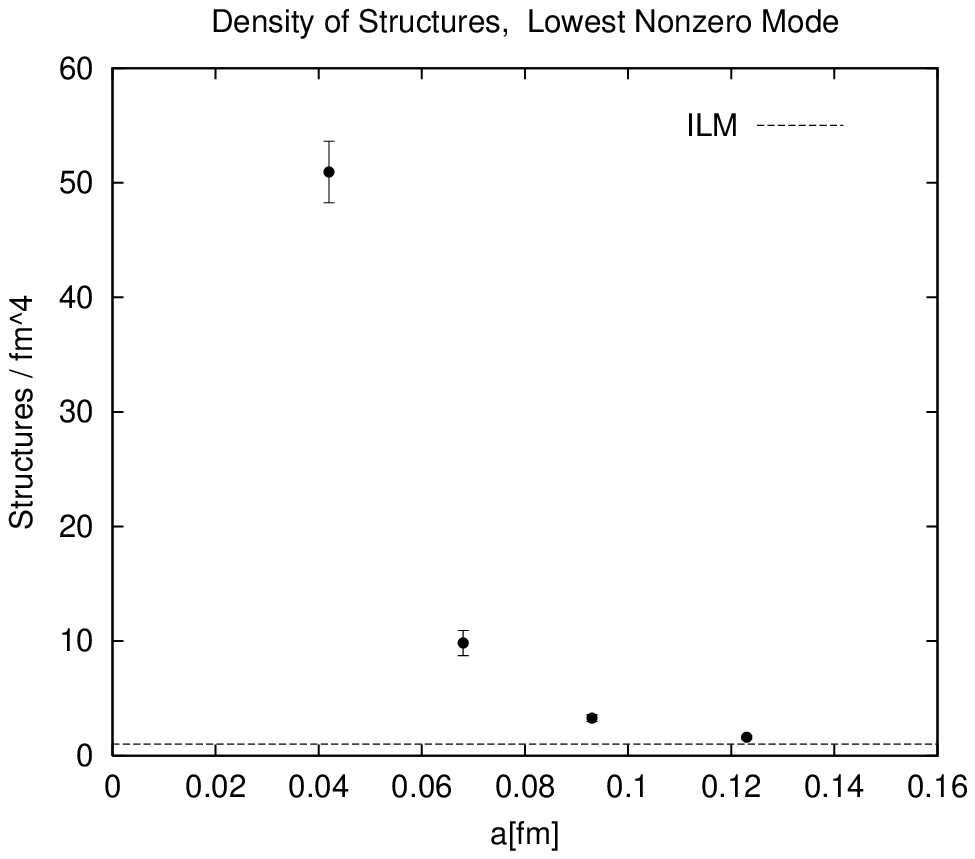}
}
\vskip 0.3truecm
\centerline{
\epsfxsize=9.0truecm\epsffile{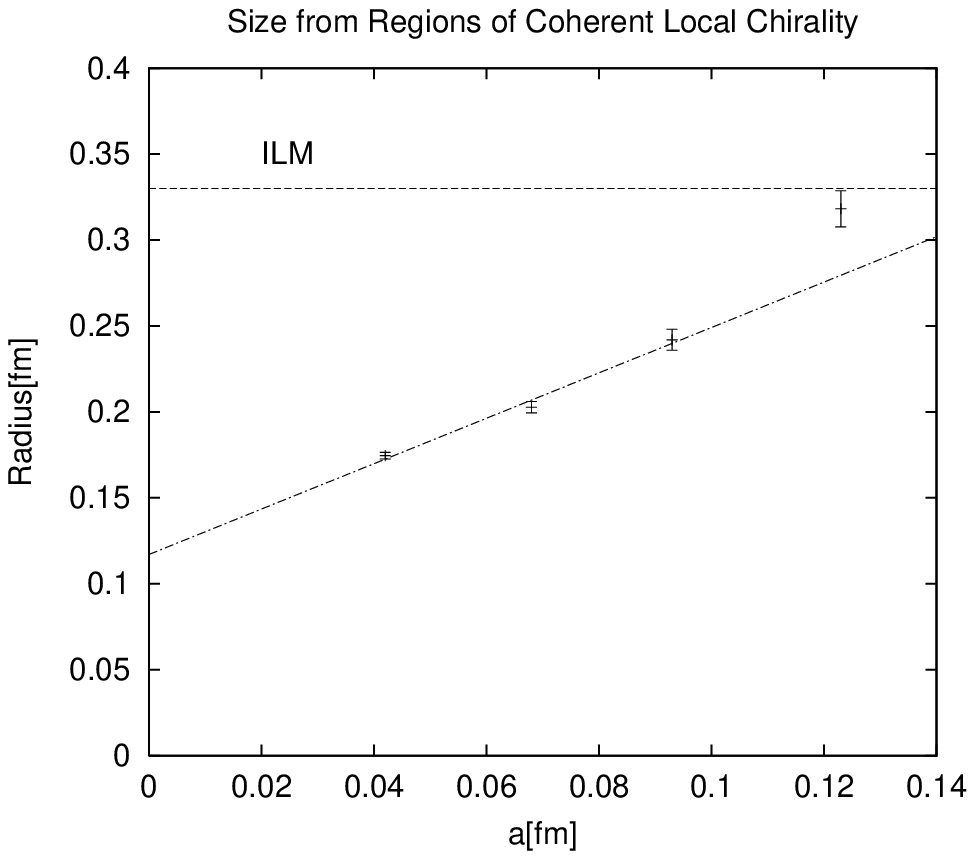}
}

\caption{Top: Density of structures (in fm$^{-4}$) as a function of the lattice 
spacing. Bottom: Average radius, $< R_n >$, of structures from regions of coherent 
local chirality. The lowest nonzero mode was used for the calculation. Data for the 
three smallest lattice spacings were used to obtain the fit. The horizontal line 
represents the radius of an ILM instanton.} 
\label{den_siz}
\end{center}
\end{figure}

As a next step we wish to establish whether the density and the size of structures
in the eigenmodes are compatible with the ILM.  In Fig.~\ref{den_siz} we show 
the lattice-spacing dependence of density indicating strikingly larger values
than assumed in the ILM. There is a rapid growth at small lattice spacings and we 
can not exclude the potential divergence in the continuum limit. This implies that 
the true local behavior of topological charge density {\it filtered} through infrared 
eigenmodes is much richer than the one pictured in the ILM. 

From our discussion in the previous section it follows that the sizes of individual 
structures can not be determined from fits to the 't Hooft profile 
(see Ref.~\cite{Hor02A} for details). We thus adopt a different procedure based 
only on the assumption of topological mixing. In particular, the ``would-be'' 
zeromodes localized on the unit quantized gauge lumps would produce regions 
of coherent local chirality of some typical size. It is thus natural to assign
the radius $R_n$ to the structure at $n$ as the radius of the largest hypersphere  
centered at $n$, containing the points with the same sign of local chirality
$\psi_m^+\gamma_5\psi_m$. We have determined the average radius for our ensembles 
and the results are shown in Fig.~\ref{den_siz}. The data shows that this quantity
has a finite continuum limit characterizing the typical size of the coherent regions.
The corresponding scale we obtain in the continuum limit is much smaller than 
the one considered in the ILM. 

\section{The Consequences of the Numerical Results}

The results from Sections 2 and 3 clearly demonstrate that the local structure 
of Dirac near-zeromodes for equilibrium QCD backgrounds differs significantly 
from the behavior associated with the ILM picture of the vacuum. This conclusion 
does not relate only to the obvious quantitative disagreement with parameters of 
the ILM that we observe. Indeed, the fact that we are not able to identify structures 
with 't Hooft profile indicates that a {\it generic} picture where vacuum is 
populated by relatively independent instantons preserving their individual identity 
is not microscopically accurate. We have thus excluded the possibility (I) of the 
Introduction.

We now consider the merit of the possibility (II), namely that quantum fluctuations,
while deforming the ILM instantons severely, do not destroy the integrity of unit 
topological charges. In such vacuum there would be identifiable individual lumps 
of unit topological charge present and the topological mixing scenario of S$\chi$SB
would still be relevant~\cite{Hor02A}. If $\chi^{j}$ is a chiral ``would-be'' zeromode
associated with a given unit lump, then the topological subspace of modes 
$\psi^i\approx \sum_j a_{ij}\chi^{j}\;,i=1,\ldots N_L$ would form if $N_L$ lumps were
present in the volume. Even for general unit lumps this situation implies very
specific properties that can be tested. 

\begin{figure}
\begin{center}
\centerline{
\epsfxsize=0.7\hsize\epsffile{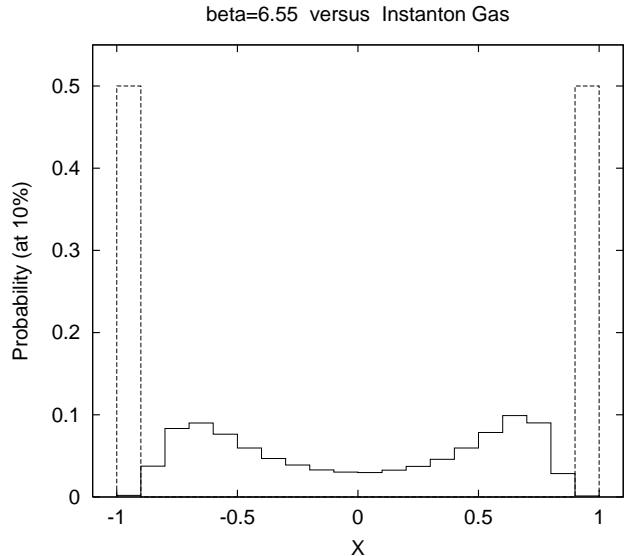}
}
\caption{Comparison of $X$-distributions at $\beta=6.55$ for near-zero modes 
(solid line) and exact zero modes (dashed line) using the overlap Dirac 
operator.} 
\label{b655bench}
\end{center}
\end{figure}

First, if a topological subspace is formed, then {\it all} eigenmodes in the subspace
should have very similar local properties since they are all the descendants 
of the chiral ``would-be'' zeromodes localized on the lumps. As a consequence, there 
should be no qualitative difference in this regard between the zeromodes and 
the near-zero modes belonging to the subspace~\cite{Hor01A}. The $X$-distribution 
is an ideal tool for testing whether this is the case or not. In Fig.~\ref{b655bench} 
we compare the $X$-distribution for two pairs of near-zeromodes from our ensemble 
at $\beta=6.55$ (our finest lattice spacing) to the $X$-distribution for the 
zeromodes~\cite{Hor02A}. Instead of qualitative agreement, we clearly observe a 
very abrupt change in the behavior of the near-zeromodes.  

\begin{figure}
\begin{center}
\centerline{
\epsfxsize=4.7truecm\epsffile{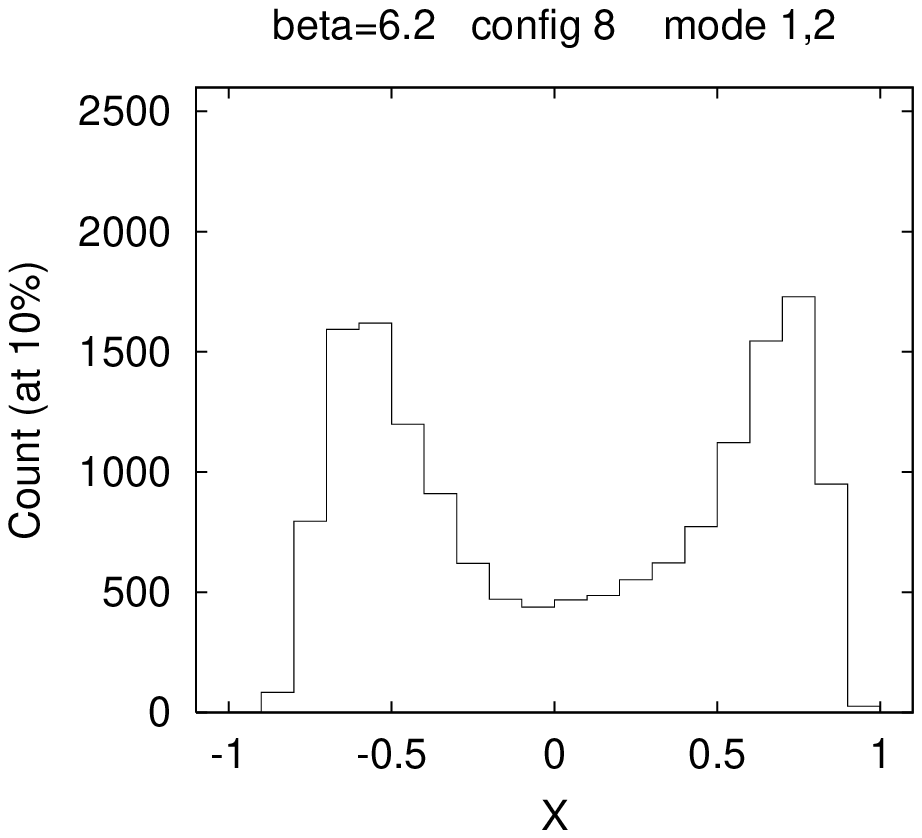}
\hspace*{-0.4truecm}
\epsfxsize=4.7truecm\epsffile{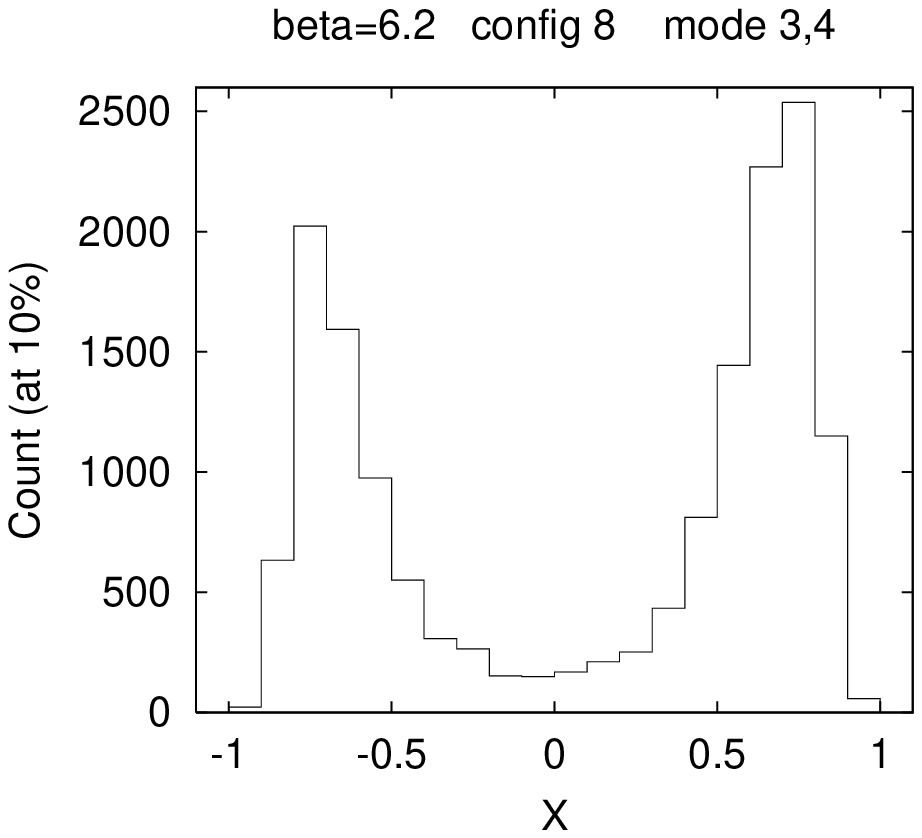}
\hspace*{-0.4truecm}
\epsfxsize=4.7truecm\epsffile{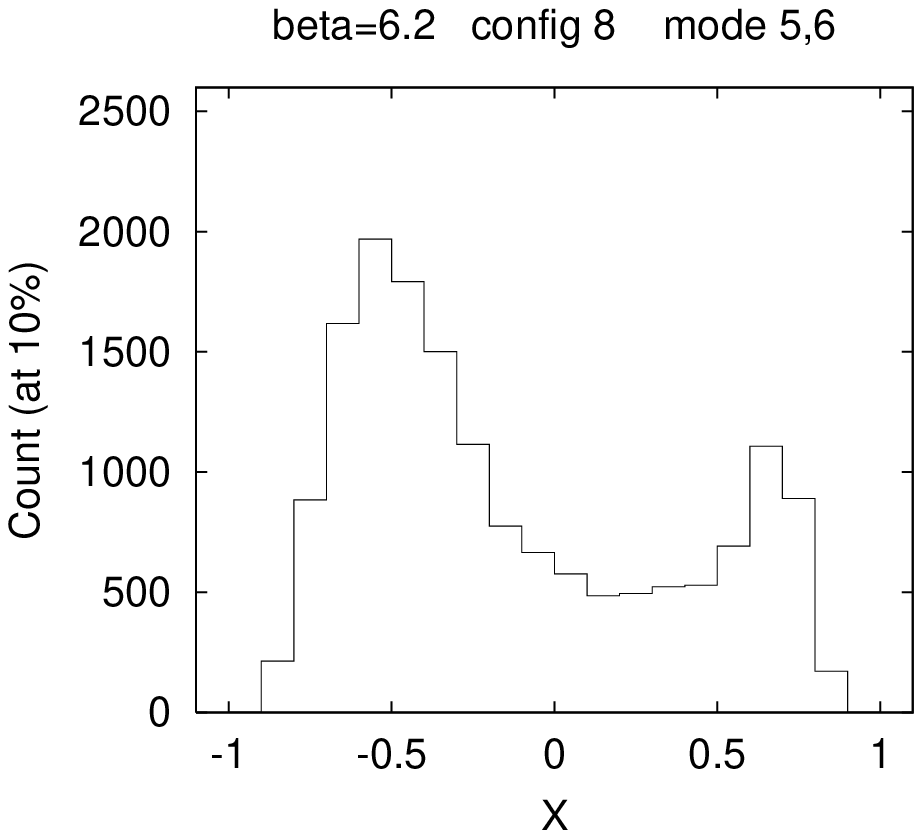}
}
\vskip 0.15in
\centerline{
\epsfxsize=4.7truecm\epsffile{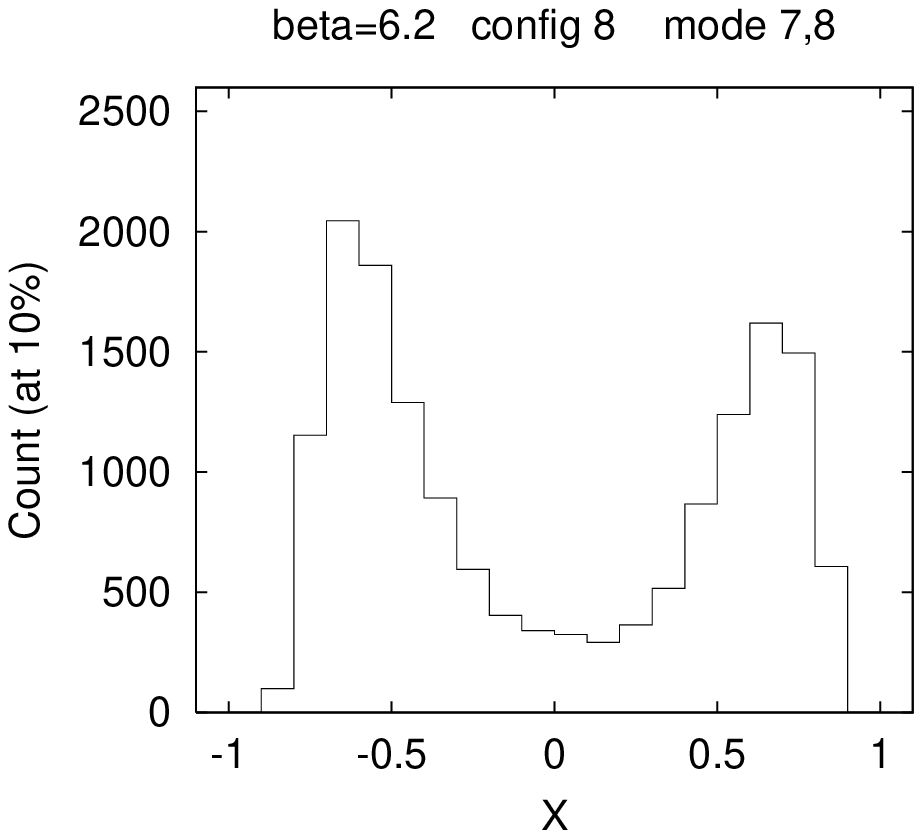}
\hspace*{-0.4truecm}
\epsfxsize=4.7truecm\epsffile{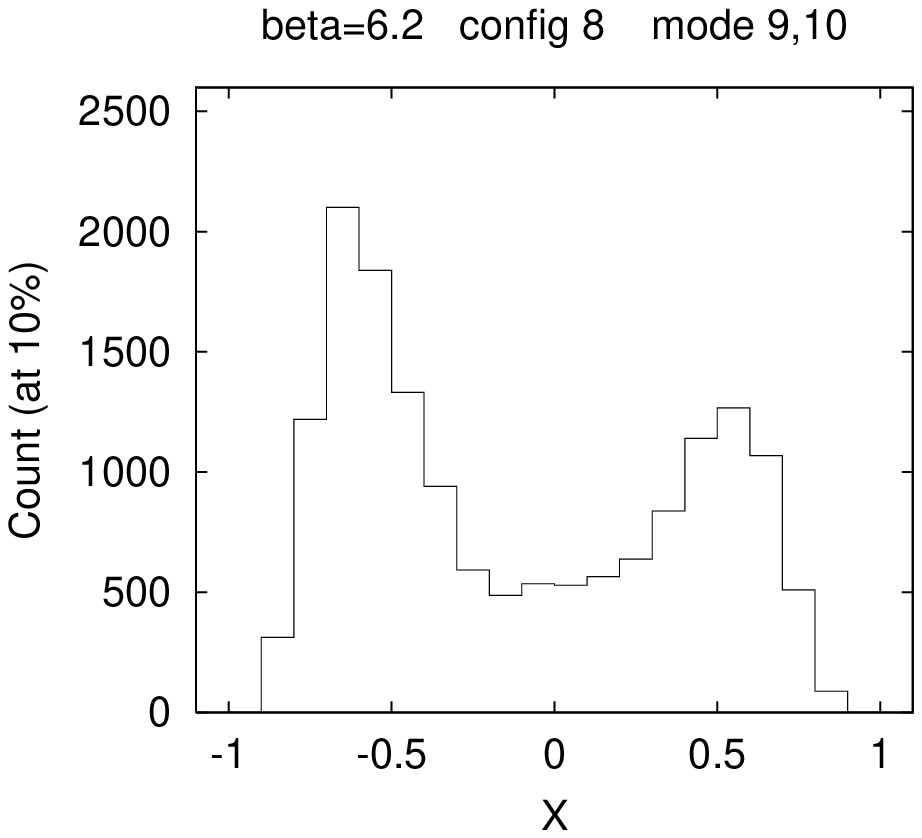}
\hspace*{-0.4truecm}
\epsfxsize=4.7truecm\epsffile{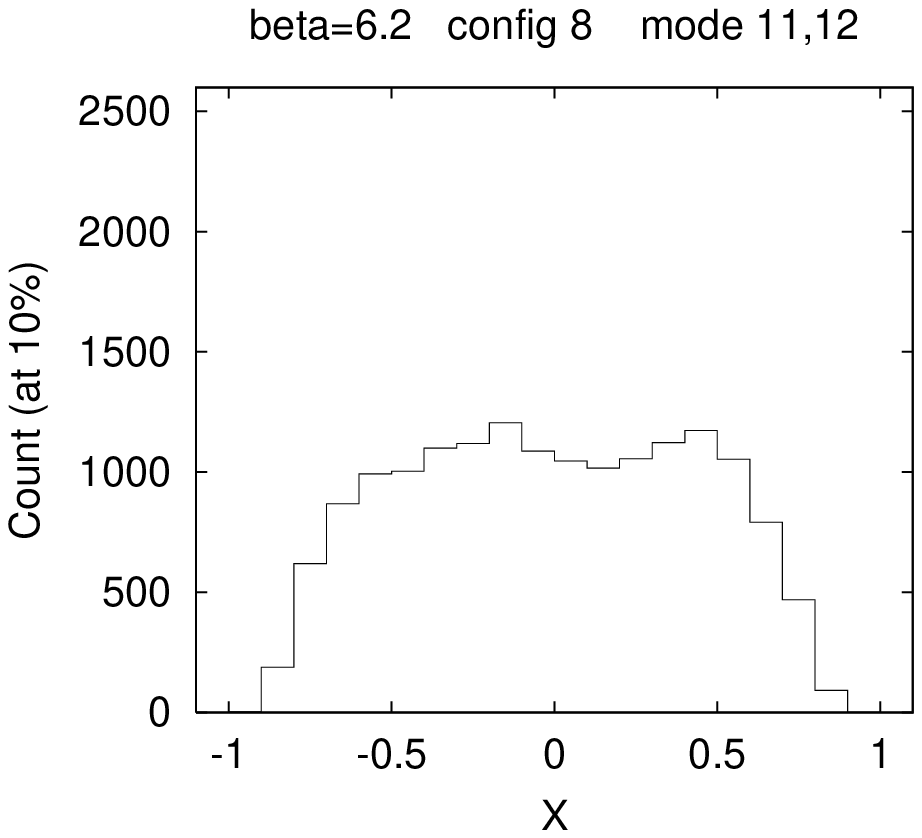}
}
\vskip 0.15in
\centerline{
\epsfxsize=4.7truecm\epsffile{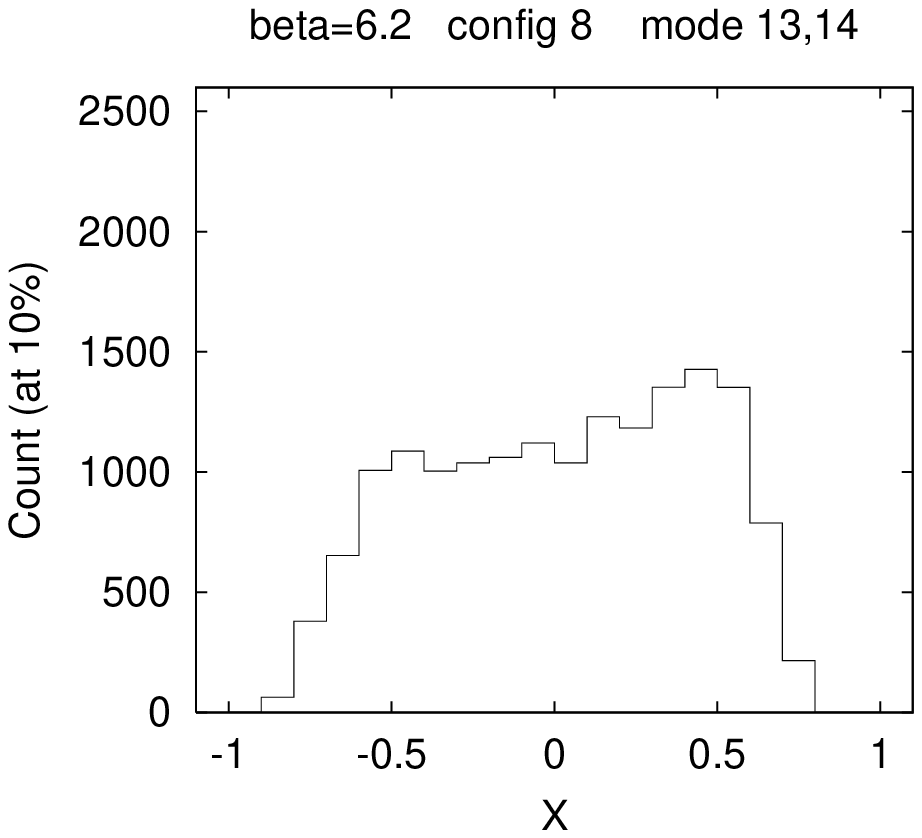}
\hspace*{-0.4truecm}
\epsfxsize=4.7truecm\epsffile{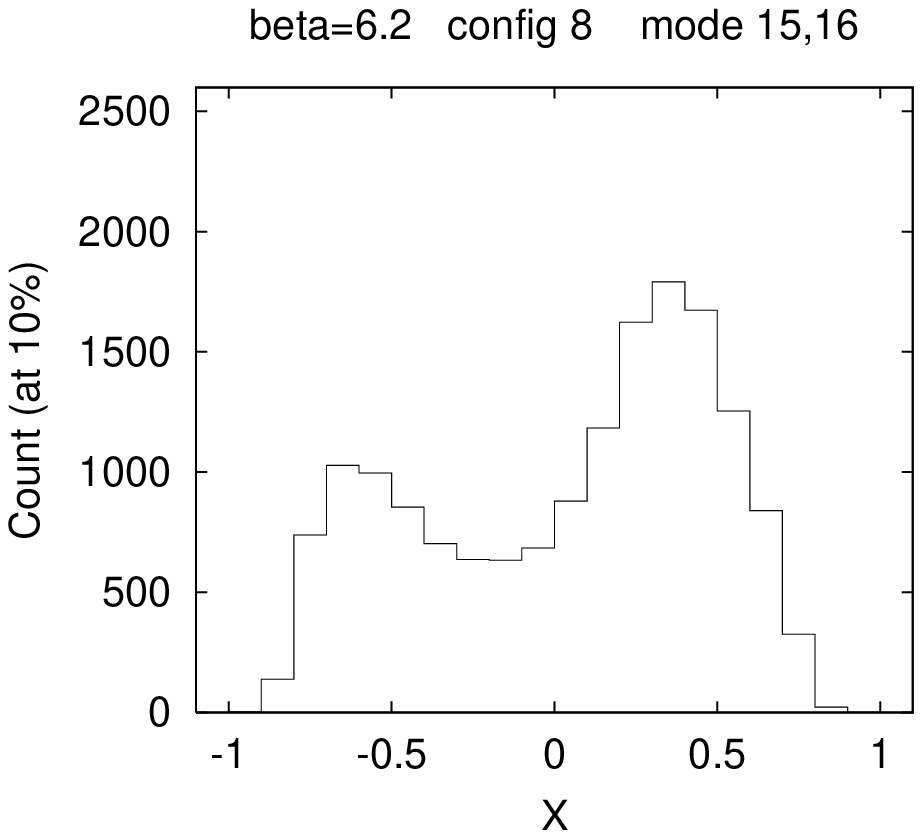}
\hspace*{-0.4truecm}
\epsfxsize=4.7truecm\epsffile{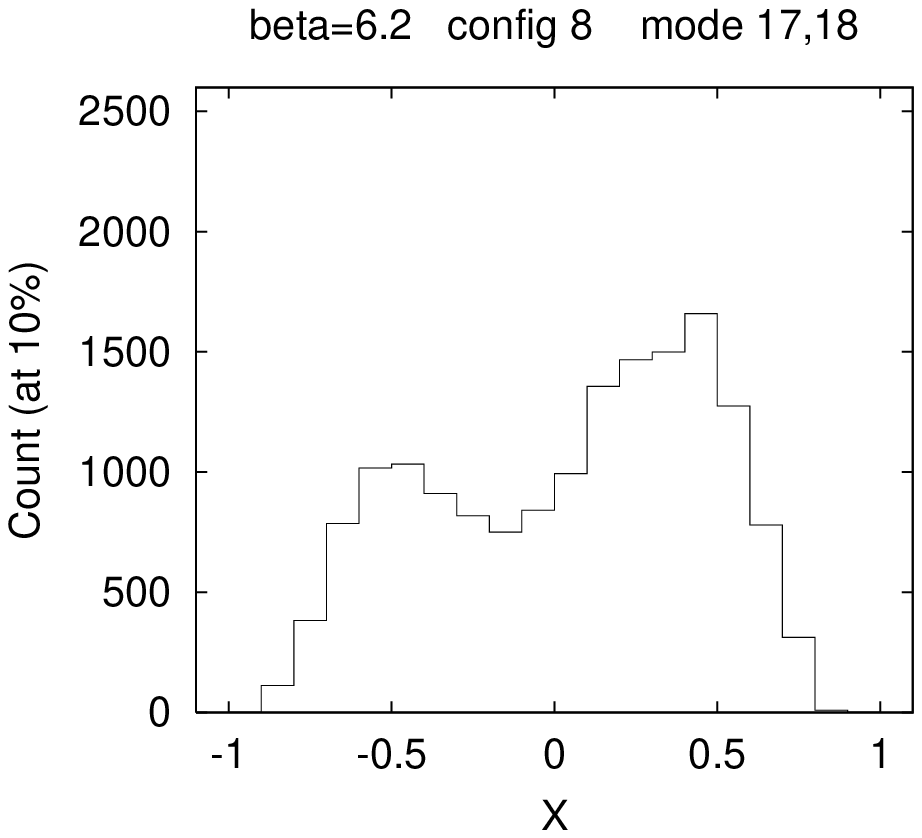}
}
\vskip 0.15in
\centerline{
\epsfxsize=4.7truecm\epsffile{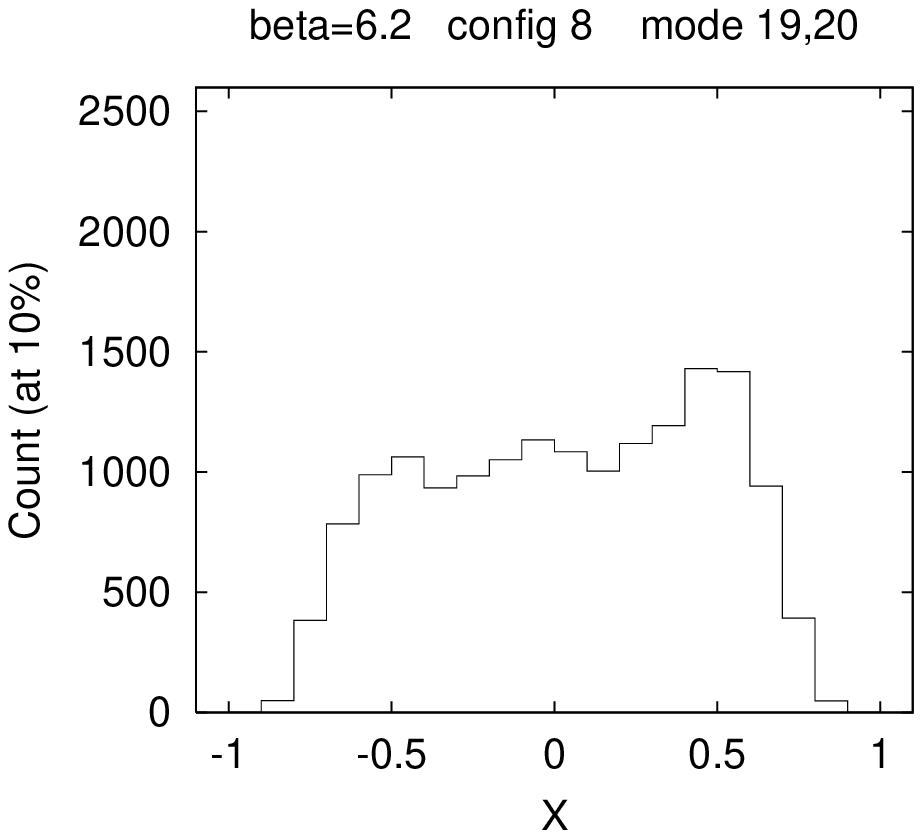}
}

\caption{The $X$-distributions for first 20 near-zero modes of configuration 8
at $\beta=6.2$.}
\label{Killer}
\end{center}
\end{figure}

At the same time, the topological mixing scenario predicts that there should be 
a sudden qualitative change in the local behavior at the point in the spectrum 
where the topological subspace ends. In particular, the $X$-distribution should be
much less peaked (or not peaked at all) for eigenmodes outside the topological 
subspace. The number of unit topological lumps $N_L$ (i.e. the dimension of the 
topological subspace) can be estimated from the known value of pure gauge topological 
susceptibility ($\approx$ 1 fm$^{-4}$). Since the lumps enter as individual entities, 
they are relatively independent, and one should observe $N_L \approx V \approx$ 3--4 
for our ensembles. For several configurations we have calculated up to $20$ 
near-zeromodes to make sure that the dimension of the topological subspace could be
identified. In Fig.~\ref{Killer} we illustrate the typical behavior by displaying 
results for configuration 8 from our $\beta=6.2$ ensemble. This configuration has 
$Q=0$ and we show the $X$-histograms for 10 pairs of near-zeromodes (the histogram 
is the same for both modes in a pair). Inspecting these results reveals that there 
are at least 14 modes with similar double-peaked structure instead of 3--4. Moreover, 
the decrease in double-peaking appears to be gradual and there is every reason 
to expect that higher modes will be peaked as well.

The above considerations show quite clearly that the idea of the distinctive topological 
subspace being the source of the low-lying eigenmodes generating the finite average 
microscopic density around zero (and hence S$\chi$SB~\cite{Banks_Casher}) is not consistent 
with our data. This suggest that the logical possibility (II) described in the Introduction 
is probably not what happens in the true vacuum of pure gauge QCD. 

\section{Conclusions}

In this work we have demonstrated that the ILM scenario for S$\chi$SB is not 
microscopically accurate. In particular, the low-lying Dirac modes cannot be described 
as mixtures of 't Hooft modes associated with ILM instantons. More generally, our data 
suggests that the bulk of topological charge in QCD is not effectively concentrated 
in quantized unit lumps to which it would be possible to assign the corresponding 
``would-be'' zeromodes. Since this talk has been given, this suggestion has been put 
on a firm ground and demonstrated directly in Ref.~\cite{Hor02B}. Our findings imply 
that a qualitatively different mechanism for the origin of the Dirac near-zeromodes 
should be sought.

\section*{Acknowledgment} 

IH thanks the organizers of the workshop for a very well-organized, informative,
and exceptionally pleasant meeting. He also acknowledges an interesting conversation 
with H.~Reinhardt and M.~Engelhardt.


\begin{thebibliography}{99}

\bibitem{ins_disc} 
   Belavin, A.A., Polyakov, A.M., Schwartz, A., Tyupkin, Y. \ (1975) \ 
   Pseudoparticle solutions of the Yang-Mills equations, 
   \textit{Physics Letters} \textbf{B59}, 85 

\bibitem{ins_gas} 
   Callan, C.G., Dashen, R., Gross, D.J. \ (1978) \  
   Toward a theory of the strong interactions,   
   \textit{Physical Review} \textbf{D17}, 2717

\bibitem{ins_tHooft} 
   't Hooft, G. \ (1976) \ 
   Symmetry breaking through Bell-Jackiw anomalies,
   \textit{Physical Review Letters}  \textbf{37}, 8;
   't Hooft, G. \ (1976) \  
   Computation of the quantum effects due to a four-dimensional pseudoparticle,
   \textit{Physical Review} \textbf{D14}, 3432

\bibitem{ins_theta} 
   Callan, C.G., Dashen, R., Gross, D.J. \ (1976) \ 
   The structure of the gauge theory vacuum, 
   \textit{Physics Letters} \textbf{B63}, 334; 
   Jackiw, R., Rebbi, C. \ (1976) \ 
   Vacuum periodicity in a Yang-Mills quantum theory,
   \textit{Physical Review Letters}  \textbf{37}, 172

\bibitem{WittenUA(1)} 
   Witten, E. \ (1979) \ 
   Instantons, the quark model, and the 1/N expansion,
   \textit{Nuclear Physics\/} \textbf{B149}, 285

\bibitem{ILM} 
   Shuryak, E.V. \ (1982) \ 
   Hadrons containing a heavy quark and QCD sum rules,
   \textit{Nuclear Physics\/} \textbf{B198}, 83; 
   Diakonov, D.I., Petrov, V.Y. \ (1984) \ 
   Instanton based vacuum from Feynman variational principle,
   \textit{Nuclear Physics\/} \textbf{B245}, 259; 
   Sch\"afer, T., Shuryak, E. \ (1998) \ 
   Instantons in QCD, 
   \textit{Reviews of Modern Physics} \textbf{70}, 323

\bibitem{ins_mixing} 
   Diakonov, D.I., Petrov, V.Y. \ (1986) \  
   A theory of light quarks in the instanton vacuum,
   \textit{Nuclear Physics\/} \textbf{B272}, 457

\bibitem{Hor02A} 
   Horv\'ath, I., et al. \ (2002) \ 
   Local chirality of low-lying Dirac eigenmodes and the Instanton
   Liquid model,
   [hep-lat/0201008]; 
   Dong, S.J., et al. \ (2002) \ 
   Topological charge fluctuations and low-lying Dirac eigenmodes,
   \textit{Nuclear Physics (Proc.\ Suppl.)}\ \textbf{106}, 563

\bibitem{Hor01A} 
   Horv\'ath, I., Isgur, N., McCune, J., Thacker, H.B. \ (2002) \ 
   Evidence against instanton dominance of topological charge fluctuations
   in QCD,
   \textit{Physical Review} \textbf{D65}, 014502

\bibitem{Hor02B} 
   Horv\'ath, I., et al. \ (2002) \ 
   On the local structure of topological charge fluctuations in QCD, 
   [hep-lat/0203027]

\bibitem{Neu98BA}
   Neuberger, H. \ (1998) \ 
   Exactly massless quarks on the lattice,
   \textit{Physics Letters} \textbf{B417}, 141;
   Neuberger, H. \ (1998) \      
   More about exactly massless quarks on the lattice,
   \textit{Physics Letters} \textbf{B427}, 353

\bibitem{Followup} 
   DeGrand, T., Hasenfratz, A. \ (2002) \ 
   Comment on ``Evidence Against Instanton Dominance of Topological Charge 
   Fluctuations in QCD'',
   \textit{Physical Review} \textbf{D65}, 014503;
   Hip, I. et al. \ (2002) \ 
   Instanton dominance of topological charge fluctuations in QCD?,
   \textit{Physical Review} \textbf{D65}, 014506;
   Edwards, R., Heller, U. \ (2002) \  
   Are topological charge fluctuations in QCD instanton dominated?,
   \textit{Physical Review} \textbf{D65}, 014505;
   Blum, T., et al. \ (2002) \ 
   Chirality Correlation within Dirac eigenvectors from Domain Wall Fermions,
   \textit{Physical Review} \textbf{D65}, 014504;
   Gattringer, C., et al. \ (2001) \
   A comprehensive picture of topological excitations in finite temperature lattice QCD,
   \textit{Nuclear Physics\/} \textbf{B618}, 205
\bibitem{Banks_Casher}
   Banks, T., Casher, A. \ (1980) \ 
   Chiral symmetry breaking in confining theories,
   \textit{Nuclear Physics\/} \textbf{B169}, 103

\end{thebibliography}
\end{document}